\documentclass[letter, 12pt]{article}

\usepackage{amsthm,amsmath,amsfonts,amssymb,amsthm,epsfig,epstopdf,titling,url,array}

\theoremstyle{plain}

\theoremstyle{definition}

\theoremstyle{remark}

\usepackage{blindtext}
\usepackage{cite}
\usepackage[utf8]{inputenc}
\usepackage{url} 
\usepackage{geometry}
\usepackage{lipsum}
\usepackage[toc,page]{appendix}
\usepackage[english]{babel}
\usepackage{graphicx} 
\usepackage{subfig}
\usepackage{authblk}

\newcommand{\threepartdef}[6]
{
	\left\{
	\begin{array}{lll}
		#1 & \mbox{if } #2 \\
		#3 & \mbox{if } #4 \\
		#5 & \mbox{if } #6
	\end{array}
	\right.
}

\usepackage{geometry}
\title{\textbf{On the distribution of cross-validated Mahalanobis distances}}

\author[1,2]{Jörn Diedrichsen}
\author[2]{Serge Provost}
\author[2]{Hossein Zareamoghaddam}
\affil[1]{\textit{Brain and Mind Institute, Western University, Canada}}
\affil[2]{\textit{Department for Statistical and Actuarial Sciences, Western University, Canada}}
\date{\today}

\begin{document}

	\maketitle
	
	\begin{abstract}
		\noindent 
		We present analytical expressions for the means and covariances of the sample distribution of the cross-validated Mahalanobis distance. This measure has proven to be especially useful in the context of representational similarity analysis (RSA) of neural activity patterns as measured by means of functional magnetic resonance imaging (fMRI). These expressions allow us to construct a normal approximation to the estimated distances, which in turn enables powerful inference on the measured statistics. Using the results, the difference between two distances can be statistically assessed, and the measured structure of the distances can be efficiently compared to predictions from computational models. 
	\end{abstract}

	\section{Introduction} 

    Representational similarity analysis (RSA) has recently gained much popularity in various branches of neuroscience \cite{RN2838, RN2697}. The main idea behind RSA is as follows: The nervous system shows complex patterns of neural activity under different conditions (for example when perceiving stimuli, performing tasks or making movements), which are a reflection of the functional processing performed by the neuronal tissue. These activity patterns can be measured using a range of different techniques: functional magnetic resonance imaging (fMRI) or magneto-encephalography (MEG) measure activity patterns on scales of mm to cm, averaging activity over millions of neurons. On the other side of the scale, electro-physiological recording of individual neurons provide a very sparse sample of the underlying population activity, but resolve the activity of the individual processing units. To understand these activity patterns, one may build models that predict the activity of each individually measured neurons or units. However, the exact spatial arrangement of the activity patterns on a fine spatial scale is rather variable across individuals \cite{RN3415}, and likely reflects random biological variation. Which neuron in a given region carries out exactly which function is to a large degree arbitrary and hence interchangeable. Thus, RSA does not analyze the distributed activity patterns themselves, but rather how different activity patterns relate to each other \cite{RN3563}. This relationship is quantified in the relative similarity or dissimilarity of two patterns. For example, if two stimuli elicit identical patterns of activity in a region, we can assume that this region is “blind” to the difference between the two conditions. If, however, different conditions evoke significantly different activity patterns, we can surmise that the region encodes something about the difference. Importantly, the pattern of relative dissimilarities across a number of different conditions can allow inferences about the latent features that underlie these neuronal representations. 
    
    RSA is especially suited to compare representations measured with different methodologies, as it does not require a 1:1 mapping between measurements. Similarly, the results can be related to the predictions of computational models, without needing to match individual measurements to individual model elements \cite{RN2838, RN2697}. In this note, we focus mostly on the application to fMRI data, for which RSA was initially developed. However, very similar techniques can be fruitfully applied to other measurement modalities as well. 
	 	
	An important factor in RSA is the choice of dissimilarity measure. In the literature, a variety of such measures have been used, including the accuracy of linear classifiers \cite{RN2348, RN3554}, Euclidean distances measures \cite{RN3563, RN2836}, or correlation-based distances \cite{RN2832, RN3560}. In this paper, we focus exclusively on one particular distance measure: The cross-validated Mahalanobis distance \cite{RN3300, RN3316}. We have recently shown that the multivariate noise normalization inherent in this measures makes it more reliable than other competing distance measures \cite{RN3540}. Furthermore, by using cross-validation, we could derive an estimate of the true Mahalanobis distance that is unbiased by the level of measurement noise, greatly improving its interpretability. The measure has also been given the name {\sl linear discriminant contrast} (LDC) \cite{RN3300} due to the close link to discriminant analysis \cite{RN3549}. 
	
	The purpose of this paper is to derive an approximation to the sample distribution of the vector of all pairwise LDCs, using the multivariate normal distribution. It is easy to show that the mean of this distribution is given by the true distance - i.e. that our estimate is unbiased. The covariance matrix of the distance estimates, however, has a more complex structure. Distances with a larger true value are estimated with higher variability. Furthermore, the estimated distance between conditions $i$ and $j$ is not independent of the estimated distances between $i$ and $k$. We derive an analytical expression for the covariance matrix and study its dependence spatial noise correlations. 
	
	These results have great utility for RSA, when making inferences based on the estimated dissimilarity measures. For example, one may wish to determine whether two activity patterns are significantly different from each other. Furthermore, one may test for linear contrasts over different distances, for example to determine if one distance is significantly larger than another. Finally, one may want to evaluate computational models using the matrix of pairwise distances. In all these cases, an approximation to the sample distribution of the distances can greatly help to improve inference, as it allows analytical expressions for the distribution under different null hypotheses, and the computation of an approximate likelihood for parameter estimation and model comparison. 
	
	In section 2, we will introduce the cross-validated Mahalanobis distance itself, focusing on the application to fMRI analysis. In section 3, we derive a multivariate normal approximation to the sampling distribution, and verify its stability under a variety of realistic assumptions in section 4. In section 5, we provide a number of practical examples of how this approximation can be fruitfully used for inference in RSA.  
	
	\section{The cross-validated Mahalanobis distance} 
	
	\subsection{Definition}
	
	In this section, we introduce the cross-validated squared Mahalanobis distance, which provides an unbiased estimate of the true squared Mahalanobis distance. 	More specifically, let $\mathbf{b}_i$  be the ``true" -- that is, noiseless -- activation pattern for the  $i^{\rm th}$ condition. In fMRI, the activation patterns are measured in terms of voxels, cubes of neural tissue of a size of approximately $1$ to $9$ mm$^3$, and dissimilarities are calculated typically for groups of $10-10000$ voxels. We define the $j^{\rm th}$ true squared Mahalanobis distance $d_j$ as that between conditions $i$ and $k$:

	\begin{eqnarray}\label{eqMahalDist}
		d_j=(\mathbf{b}_i-\mathbf{b}_k)\boldsymbol{\Sigma}_P^{-1}(\mathbf{b}_i-\mathbf{b}_k)^T/P, 
	\end{eqnarray}
		
	\noindent where $\boldsymbol{\Sigma}_P$ is the voxel-by-voxel noise-covariance matrix. Furthermore, we have normalized the distances by the number of voxels, denoted by $P$, to make the measure comparable across regions of different sizes. For $K$ conditions, we would have a total of $K(K-1)/2\equiv D$ unique distances. 
	
	\subsection{Estimation of activity patterns}
	
	Let $\mathbf{B}$ be the $K\times P$ matrix of true activation patterns. To estimate these patterns, we present each condition multiple times to the participant, each time measuring the neuronal response. In the context of fMRI, the data comes in terms of time-series of blood-oxygenation level dependent (BOLD) signal measurements. The time series is usually separated into $M$ different runs or phases of data recording. Because the responses in fMRI are temporally extended (up to $30$s), we obtain activation estimates by deconvolving the BOLD signal through a  first-level linear model analysis \cite{RN2030}. In this approach, the time series from each run is modeled using the design matrix $\mathbf{X}_m$, which has one column per condition. For each condition, the phases of activity are convolved with an estimate of the hemodynamic response function \cite{RN2028} to yield a predicted response. The time series data are then modeled as the product of a design matrix and the true activity patterns:

	\begin{eqnarray}\label{eqEstModel}
	\mathbf{Y}_m=\mathbf{X}_m\,\mathbf{B}_m+\boldsymbol{\epsilon}_m, 
	\end{eqnarray}	
	\noindent where the random errors $\boldsymbol{\epsilon}_m$ have a temporal autocorrelation \cite{RN2030, RN2173}, and possibly non-stationary variance along the time dimension \cite{RN2372}. After determining an estimate of the covariance matrix $\boldsymbol{\Sigma}_T$ along the time axis, we can obtain an optimal general least-squares estimate of the activity patterns from the $m^{\rm th}$ run

	\begin{eqnarray}\label{eqEstBeta}
	{\hat{\mathbf{B}}}_m=(\mathbf{X}_m^T\hat{\boldsymbol{\Sigma}}_T^{-1}\mathbf{X}_m)^{-1}\mathbf{X}_m^{T}\hat{\boldsymbol{\Sigma}}_T^{-1}{\mathbf{Y}}_m.
	\end{eqnarray}
	
	We assume that the activity estimates are independent across runs. However, because of the temporal autocorrelation of the noise and the slow hemodynamic response, which leads to overlapping responses from different conditions, we generally cannot assume independence of activation estimates within a single run. 
	
	\subsection{Spatial pre-whitening of activity patterns}
	
	 We are interested in the Mahalanobis distance, rather than the simpler Euclidean distance, for two reasons: First, multivariate noise-normalization (weighting the quadratic form by $\boldsymbol{\Sigma}_P^{-1}$) leads to more reliable distance estimates than univariate noise normalization, i.e. dividing by the diagonal entries of $\boldsymbol{\Sigma}_P$ \cite{RN3540}. Secondly, we may want to give less weight to the information contained in two voxels or neurons that are highly correlated in their random variability than to information contained in two uncorrelated neurons. In short, the Mahalanobis distance measures the amount of information extracted by an optimal Gaussian linear decoder. 
	
	In practice, we do not have access to the voxel-by-voxel covariance matrix. However, we can use the residuals $\mathbf{R}_m$ from the regression analysis (Eq. \ref{eqEstBeta}) to derive an estimate, 
	
	\begin{eqnarray}\label{eqWhiteSigmaP}
	\hat{\boldsymbol{\Sigma}}_P=\frac{1}{M(T_m-K-Q)}\sum_{m=1}^{M}\mathbf{R}_m^T\mathbf{R}_m,
	\end{eqnarray}	
	
	\noindent where $K$ is the number of regressors of interest and $Q$ the number nuisance regressors in each $\mathbf{X}_m$. Oftentimes, we have the case that $P>T$, which renders the estimate non-invertible. Even with $T>P$, it is usually prudent to regularize the estimate, as it stabilizes the distance estimates. A practical way of doing this is to shrink the estimate of the covariance matrix to a diagonal version of the raw sample estimate:

	\begin{eqnarray}\label{eqWhiteRegular}
	\tilde{\boldsymbol{\Sigma}}_P=h\,\text{diag}(\hat{\boldsymbol{\Sigma}}_P)+(1-h)\,\hat{\boldsymbol{\Sigma}}_P.
	\end{eqnarray}	
	
	\noindent The scalar $h$ determines the amount of shrinkage, with $0$ corresponding to no shrinkage and $1$ to only using the diagonal of the matrix. Estimation methods for the optimal shrinkage coefficient have been proposed \cite{RN3119}, but usually values around $h=0.4$ perform well for fMRI data analysis (see below). 

	This estimate is used for determining Mahalanobis distances. For convenience of notation, we can operate on spatially prewhitened versions of $\hat{\mathbf{B}}_m$:

	\begin{eqnarray}\label{eqWhitePrewhite}
	\hat{\mathbf{U}}_m=\hat{\mathbf{B}}_m\tilde{\boldsymbol{\Sigma}}_P^{-1/2}.
	\end{eqnarray}			
	
	\noindent Using this notation, the Mahalanobis distance simplifies to the Euclidean distance between prewhitened activity patterns. 
	
	To derive the statistical properties of distances calculated on the basis of these activity estimates, we shall assume that $\hat{\mathbf{U}}_m$ has matrix-variate normal distribution \cite{RN3554}:

	\begin{eqnarray}\label{eqWhiteMN}
	\hat{\mathbf{U}}_{m} \sim\mathcal{MN} (\mathbf{U},\mathbf{\Sigma}_{K},\mathbf{\Sigma}_{R}). 
	\end{eqnarray}	
	
	The mean centers on the true (prewhitened) activity patterns $\mathbf{U}$. The covariance between rows (conditions) is $\boldsymbol{\Sigma}_K$, which captures the variance of (and covariance between) the activity estimates from the same partition. Finally, the columns (voxels) are assumed to have a residual covariance $\boldsymbol{\Sigma}_R$ that persists even after prewhitening. This residual covariance arises from the fact that the regularised estimate of the covariance matrix (Eq. \ref{eqWhiteRegular}) will differ somewhat from the true voxel-by-voxel covariance matrix $\boldsymbol{\Sigma}_P$ :

	\begin{eqnarray}\label{eqWhiteSigmaR}
		{\boldsymbol{\Sigma}}_R=\tilde{\boldsymbol{\Sigma}}_P^{-1/2}\boldsymbol{\Sigma}_P\tilde{\boldsymbol{\Sigma}}_P^{-1/2}.
	\end{eqnarray}

	\subsection{Cross-validation}
	
	One way to calculate pattern distances is to simply substitute the activity patterns in Eq. \ref{eqMahalDist} with their noisy estimates. However, because the noise enters into a quadratic form, it is squared and causes a positive bias. Specifically, distances on estimated activity patterns will always be larger than zero, even if the true value is zero. In general, the expected value of the distance estimate will increase with increasing noise level. This complicates the comparison of different distances (and especially their ratios) between different brain regions or individuals, which can differ considerably in the level of measurement noise. 
	
	We can obtain an unbiased estimate for the pattern distances using cross-validation.  We first define $\boldsymbol{\hat{\delta}}_j,m$ to be the difference between two corresponding activity pattern estimates from the $m^{\rm th}$ partition as

	\begin{eqnarray}\label{eqCrossDiff}
	\hat{\boldsymbol{\delta}}_{j,m}=\hat{\mathbf{u}}_{i,m}-\hat{\mathbf{u}}_{k,m}
	\end{eqnarray}			
	
	\noindent and then calculate the distance estimate as the inner product between pairs of independent estimates of $\boldsymbol{\delta}_j$:

	\begin{eqnarray}\label{eqCrossDist}
	\hat{d}_{j}=\frac{1}{MP}\sum_{i=1}^{m}\hat{\boldsymbol{\delta}}_{j,m}\hat{\boldsymbol{\delta}}_{j,\sim m}^T,
	\end{eqnarray}	
	
	\noindent where $\hat{\boldsymbol{\delta}}_{j,m}$ is the estimate of the pattern difference based on partition $m$, and $\hat{\boldsymbol{\delta}}_{j,\sim m}$  is the estimate of the pattern difference estimated from all partitions except $m$. The latter estimate is usually obtained by averaging across all partitions except $m$ (however, see sections 3.4-3.5 for more details).  
	
	\section{Sample distribution of LDC distances}
	
	\subsection{Normal approximation to the joint sample distribution of LDC distances}
	
	In this section, we derive an approximation to the sample distribution of LDC distances. Specifically, we will show that the vector of all possible pairwise distances $\hat{\mathbf{d}}$ has approximately a multivariate normal distribution, with  
	
	\begin{eqnarray}\label{eqApproxNorm}
	\hat{\mathbf{d}} \sim\mathcal{N}_D (\mathbf{d},\,\mathbf{V}). 
	\end{eqnarray}		
	
	\noindent where the mean is given by the vector of true distances $\mathbf{d}$. The covariance matrix $\mathbf{V}$ depends on the true distances and the covariance structure of the activity estimates. 
	
	It can be relatively easily seen that the mean of the sampling distribution is the vector of true distances $\mathbf{d}$, i.e., that the estimate is unbiased. Consider the estimate of each pattern difference to be the true pattern difference plus independent noise, i.e., $\hat{\boldsymbol{\delta}}_{j, m} = \boldsymbol{\delta}_{j, m} + \boldsymbol{\epsilon}_{j, m}$, where $\boldsymbol{\epsilon}$ is both independent of the true difference and of the noise terms of other partitions. Then,
	
	\begin{align}\label{eqApproxMean}
	E(\hat{d}_{j}) 
	=&\,\, E \Big(\frac{1}{MP}\sum_{i=1}^{m}(\boldsymbol{\delta}_{j} + \boldsymbol{\eta}_{j, m})(\boldsymbol{\delta}_{j} + \boldsymbol{\eta}_{j, \sim m})^T\Big) \nonumber \\
	=&\,\, E \Big(\frac{1}{MP}\sum_{i=1}^{m}\boldsymbol{\delta}_{j}\boldsymbol{\delta}_{j}^T + 
	\boldsymbol{\delta}_{j}\boldsymbol{\eta}_{j, m}^T + 
	\boldsymbol{\delta}_{j}\boldsymbol{\eta}_{j, \sim m}^T + 
	\boldsymbol{\eta}_{j, m} \boldsymbol{\eta}_{j, \sim m}^T\Big)	\nonumber \\
	=&\,\, \boldsymbol{\delta}_{j}\boldsymbol{\delta}_{j}^T/P = d_{j}.
	\end{align}

\begin{figure}
		\centering
		\includegraphics[width=14cm]{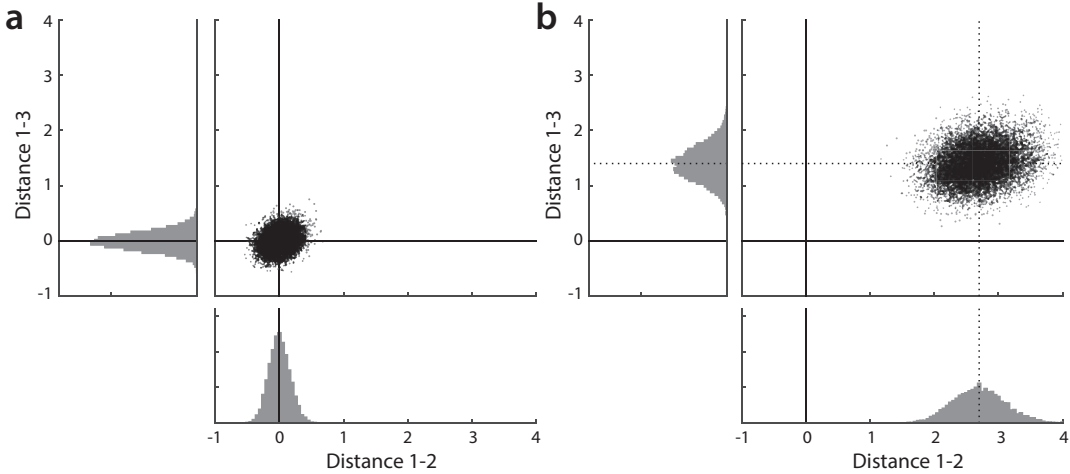}
		\caption{Joint sample distribution of the distance estimates (LDC) between conditions 1 and 2 (x-axis) and conditions 1 and 3 (y-axis). Samples are based on simulated data with $P=50$ voxels, $M = 3$ partitions and $K = 3$ conditions. The main panel shows the joint scatter plot of the samples, the lateral insets showing the marginal distributions. (\textbf{a}) Sample distributions for the case wherein the true distances $d_{12}$ and $d_{13}$ are zero. (\textbf{b}) Sample distributions when the true distances are set to $d_{1,2}=2.6$ and $d_{1,3}=1.4$ (dotted lines). }
		\label{fig1}
	\end{figure}

	A simulation example of the sample distribution of the LDC is shown in Figure \ref{fig1}. The simulated example includes a modest number of voxels ($50$) and a low number of partitions. Despite this, it is clear that the joint distribution  of the two distances can be well approximated by a multivariate normal, and that the distribution centers on the true value of the simulated distance (dotted line).

   The covariance matrix of the vector distances, however, has more complex features, which are clearly apparent in the simulation. First, the variability of the distance estimates increases as the true distances increase (Fig. \ref{fig1}b) - that is, the variance depends on the mean. Furthermore, the distance estimate between conditions 1 and 2 and the distance estimate between conditions 1 and 3 are not independent, but positively correlated. In the remainder of this section, we will show that the covariance of the distance estimates depends on ${\mathbf{d}}$, the vector of true distances, $\mathbf{\Sigma}_{K}$, the covariance of the pattern estimates across rows or conditions, and $\mathbf{\Sigma}_{R}$, the residual covariance of the pattern estimates across columns or voxels, namely that

	\begin{eqnarray}\label{eqApproxVar}
	\mathbf{V} =\Big[4\frac{\boldsymbol{\Delta}  \circ \boldsymbol{\Xi}}{M} + 2\frac{\boldsymbol{\Xi}  \circ \boldsymbol{\Xi}}{M(M-1)}  \Big]  \frac{{\rm tr}(\boldsymbol{\Sigma}_R\boldsymbol{\Sigma}_R)}{P^2},
	\end{eqnarray}		
	
	\noindent where ${\circ}$ denotes the element-by-element multiplication of two matrices. 
	The matrix $\boldsymbol{\Delta}$ is the second-moment matrix of the activity differences of the true activation patterns, with 
	\begin{eqnarray}\label{eqxx}
		{\Delta}_{i,k}=\boldsymbol{\delta}_i \boldsymbol{\delta}_k^T/P.
	\end{eqnarray}	
	To express this quantity in matrix notation, we can define a $D \times K$ contrast matrix $\mathbf{C}$. The $j^{th}$ row of this matrix contains a $1$ and a $-1$ for the two conditions that are contrasted in the $j^{th}$ distance. $\boldsymbol{\Delta}$ can then be simply expressed as

	\begin{eqnarray}\label{eqApproxDelta}
	\boldsymbol{\Delta}= \mathbf{C}\mathbf{U}\,\mathbf{U}^T\mathbf{C}^T /P =-\frac{1}{2}\mathbf{C}\,\mathbf{D}\,\mathbf{C}^T.
	\end{eqnarray}

	\noindent The last equality implies that we do not need the full second-moment matrix of the true activity patterns, $\mathbf{U}\,\mathbf{U}^T/P$, but only the vector of true distances $\mathbf{d}$, arranged in a $K\times K$ representational dissimilarity matrix $\mathbf{D}$. 
	Finally, $\boldsymbol{\Xi}$ is the covariance matrix of the estimated pattern differences across the different partitions, and depends simply on the covarariance of the rows of $\hat{\mathbf{U}}_{m}$:

	\begin{eqnarray}\label{eqApproxXi}
	\boldsymbol{\Xi}={\rm Cov}(\mathbf{C}\,\hat{\mathbf{U}}_m)=\mathbf{C}\boldsymbol{\Sigma}_K\mathbf{C}^T.
	\end{eqnarray}	
	
	\noindent
	From Eq. \ref{eqApproxVar}, it is apparent that the (co-)variance of the distance estimates contains two components: a signal-dependent part that relies on the “true” distances and a signal-independent part that only depends on the variability of the activity estimates across different partitions. 
	
	In the next sections, we will derive the representation of $\mathbf{V}$ given in Eq. \ref{eqApproxVar}, including a more general form thereof, by making use of basic statistical results on the inner product of two random vectors. 
	
	\subsection{Statistical properties of inner products – independent elements}
	
	In this section, we will assume that the distinct elements (voxels) of the random vectors are independent of each other. We will generalize these results to vectors with correlation between voxels in section 3.3. For now, suppose we have two $1\times P$ row random vectors
	
	\begin{eqnarray} \label{eqProd1}
	\hat{\boldsymbol{\delta}}_A = \boldsymbol{\delta}_A+\boldsymbol{\epsilon}_A \nonumber \\
		\hat{\boldsymbol{\delta}}_B = \boldsymbol{\delta}_B+\boldsymbol{\epsilon}_B
	\end{eqnarray}

	\noindent where ${\boldsymbol{\delta}}_A$ and ${\boldsymbol{\delta}}_B$ are constant $1\times P$ vectors, and $\boldsymbol{\epsilon}_A$  and $\boldsymbol{\epsilon}_B$ being vectors of independent random variables with mean zero and variances $\sigma_A^2$ and $\sigma_B^2$, respectively. We are initially interested in the expected value and the variance of the inner product $\hat{\boldsymbol{\delta}}_A\hat{\boldsymbol{\delta}}^T_B$, which can be expressed as follows
	
	\begin{eqnarray}\label{eqProd2}
	\hat{\boldsymbol{\delta}}_A\hat{\boldsymbol{\delta}}^T_B={\boldsymbol{\delta}}_A{\boldsymbol{\delta}}^T_B+{\boldsymbol{\delta}}_A\boldsymbol{\epsilon}_B^T+\boldsymbol{\epsilon}_A {\boldsymbol{\delta}}^T_B+\boldsymbol{\epsilon}_A\boldsymbol{\epsilon}_B^T.
	\end{eqnarray}		
	
	\noindent Because the expected inner product between the two independent noise vectors, and between a noise and a constant vector are zero, we have

	\begin{eqnarray}\label{eqProdMean}
	E\Big( \hat{\boldsymbol{\delta}}_A\hat{\boldsymbol{\delta}}^T_B \Big)={\boldsymbol{\delta}}_A{\boldsymbol{\delta}}^T_B.
	\end{eqnarray}

   In order to determine the variance of the inner product, we first consider the variance of the product of two independent random variables. Making use of the identity
	
	\begin{eqnarray}\label{eqProdVar1}
	{\rm var}(xy)=E(x)^2{\rm var}(y)+E(y)^2{\rm var}(x)+{\rm var}(x){\rm var}(y),
	\end{eqnarray}	
	
	\noindent we obtain the following result for the inner product of two independent random vectors: 
	
	\begin{align}\label{eqProdVar2}
	{\rm var}(\hat{\boldsymbol{\delta}}_A\hat{\boldsymbol{\delta}}^T_B)&={\rm var}({\boldsymbol{\delta}}_A{\boldsymbol{\delta}}^T_B+{\boldsymbol{\delta}}_A\boldsymbol{\epsilon}_B^T+\boldsymbol{\epsilon}_A {\boldsymbol{\delta}}^T_B+\boldsymbol{\epsilon}_A\boldsymbol{\epsilon}_B^T) \nonumber \\
	&=0+{\boldsymbol{\delta}}_A{\boldsymbol{\delta}}^T_A\sigma^2_B+{\boldsymbol{\delta}}_B{\boldsymbol{\delta}}^T_B\sigma^2_A+P\sigma^2_A\sigma^2_B.
	\end{align}

	Finally, we need an expression for the covariance between two inner products. As before, we assume that the random vectors $\hat{\boldsymbol{\delta}}_A$ and $\hat{\boldsymbol{\delta}}_B$ are independent of each other, and that   $\hat{\boldsymbol{\delta}}_C$ and $\hat{\boldsymbol{\delta}}_D$  are also mutually independent; however, we allow the elements of the other pairs of vectors to share covariance  $\gamma_{\cdot,\,\cdot}$. Thus,

		\begin{align}\label{eqProdCov}
		{\rm Cov}(\hat{\boldsymbol{\delta}}_A\hat{\boldsymbol{\delta}}_B^T,\,\hat{\boldsymbol{\delta}}_C\hat{\boldsymbol{\delta}}_D^T)=&\,\, {\rm Cov}\Big[ \big({\boldsymbol{\delta}}_A+\boldsymbol{\epsilon}_A\big)\big({\boldsymbol{\delta}}_B+\boldsymbol{\epsilon}_B\big)^T,\, \big({\boldsymbol{\delta}}_C+\boldsymbol{\epsilon}_C\big)\big({\boldsymbol{\delta}}_D+\boldsymbol{\epsilon}_D\big)^T  \Big] \nonumber \\
		=&\,\, {\rm Cov}\Big[ {\boldsymbol{\delta}}_A{\boldsymbol{\delta}}^T_B+{\boldsymbol{\delta}}_A\boldsymbol{\epsilon}_B^T+\boldsymbol{\epsilon}_A {\boldsymbol{\delta}}^T_B+\boldsymbol{\epsilon}_A\boldsymbol{\epsilon}_B^T, \,
		{\boldsymbol{\delta}}_C{\boldsymbol{\delta}}^T_D+{\boldsymbol{\delta}}_C\boldsymbol{\epsilon}_D^T+\boldsymbol{\epsilon}_C {\boldsymbol{\delta}}^T_D+\boldsymbol{\epsilon}_C\boldsymbol{\epsilon}_D^T \Big] \nonumber \\
		=&\,\, {\boldsymbol{\delta}}_A{\boldsymbol{\delta}}^T_C\, \gamma_{B,D}+{\boldsymbol{\delta}}_A{\boldsymbol{\delta}}^T_D\, \gamma_{B,C}+{\boldsymbol{\delta}}_B{\boldsymbol{\delta}}^T_C\,  \gamma_{A,D}+{\boldsymbol{\delta}}_B{\boldsymbol{\delta}}^T_D\, \gamma_{A,C}\nonumber \\
		&\,\, +P\big( \gamma_{A,C}\gamma_{B,D}+\gamma_{A,D}\gamma_{B,C} \big).
		\end{align}

\noindent When $A=C$ and $B=D$, this simplifies to Eq. \ref{eqProdVar2}.

\subsection{Dependence across voxels}

The previous result on products between random variables needs to be extended to the situation wherein the elements of the random vector are not mutually independent, but have a known covariance structure. Such a dependence structure arises in the context of imaging: Spatial noise correlations are ubiqitous, and prewhitening is usually incomplete, leaving a remaining dependence between voxels. 

Analogous to Eq. \ref{eqProdCov}, let us consider the covariance between two products of activity estimates from two different voxels $i$ and $j$. The covariance of the activation estimates of these two voxels within partition A is denoted by $\sigma_{A,i,j}$. The two activity estimates that are multiplied are always independent, as they come from non-overlapping partitions. We can therefore apply the result from Eq. \ref{eqProdCov} to obtain

	\begin{eqnarray}\label{eqSpatialCov1}
	{\rm Cov}(\hat{\delta}_{A,i} \hat{\delta}_{B,i}, \hat{\delta}_{A,j}\hat{\delta}_{B,j}) 
	=\delta_{B,i}\delta_{B,j}\sigma_{A,i,j}+\delta_{A,i}\delta_{A,j}\sigma_{B,i,j}+\sigma_{A,i,j}\sigma_{B,i,j}.
	\end{eqnarray}	

\noindent Thus, the full covariance matrix of the element-wise products ($\circ$) between two random vectors can be written as 
	\begin{eqnarray}\label{eqSpatialCov2}
	{\rm Cov}(\hat{\boldsymbol{\delta}}_{A}\circ\hat{\boldsymbol{\delta}}_{B})=\boldsymbol{\delta}_{B}^T\boldsymbol{\delta}_{B}\circ\boldsymbol{\Sigma}_{A}+\boldsymbol{\delta}_{A}^T\boldsymbol{\delta}_{A}\circ\boldsymbol{\Sigma}_{B}+\boldsymbol{\Sigma}_{A}\circ\boldsymbol{\Sigma}_{B}
	\end{eqnarray}

\noindent where for instance $\boldsymbol{\Sigma}_{A}$ is the covariance matrix across voxels for partition A. The inner product is now simply the sum of the element-wise products:  

\begin{eqnarray}\label{eqSpatialCov3}
\hat{\boldsymbol{\delta}}_{A}\hat{\boldsymbol{\delta}}_{B}^T=\mathbf{1}^T\big(\hat{\boldsymbol{\delta}}_{A}\circ\hat{\boldsymbol{\delta}}_{B} \big)
\end{eqnarray}		

\noindent where $\mathbf{1}$ is a column vector of ones. Consequently, the variance of the inner product can be expressed as follows:

\begin{align}\label{eqSpatialCov4}
{\rm var}\big( \hat{\boldsymbol{\delta}}_{A}\hat{\boldsymbol{\delta}}_{B}^T \big)&= \mathbf{1}^T {\rm Cov}\big(\hat{\boldsymbol{\delta}}_{A}\circ\hat{\boldsymbol{\delta}}_{B} \big)\mathbf{1}   \nonumber \\
&=\mathbf{1}^T \big(\boldsymbol{\delta}_{B}^T\boldsymbol{\delta}_{B}\circ\boldsymbol{\Sigma}_{A}+\boldsymbol{\delta}_{A}^T\boldsymbol{\delta}_{A}\circ\boldsymbol{\Sigma}_{B}+\boldsymbol{\Sigma}_{A}\circ\boldsymbol{\Sigma}_{B} \big)\mathbf{1}   \nonumber \\
&={\rm tr}\Big( \boldsymbol{\delta}_{B}^T\boldsymbol{\delta}_{B}\boldsymbol{\Sigma}_{A}+\boldsymbol{\delta}_{A}^T\boldsymbol{\delta}_{A}\boldsymbol{\Sigma}_{B}+\boldsymbol{\Sigma}_{A}\boldsymbol{\Sigma}_{B} \Big).
\end{align}	

To progress further, we now have to make the simplifying assumption that the spatial structure of the true mean signal ($\boldsymbol{\delta}$) and the spatial structure of the random variation are the same, that is, that the expected outer products are equal to the matrix $\boldsymbol{\Sigma}_{R}$ scaled by different constants:

\begin{align}\label{eqSpatialCov5}
\boldsymbol{\Sigma}_A&=\boldsymbol{\Sigma}_R\sigma_A^2, \nonumber \\ 		
\boldsymbol{\Sigma}_B&=\boldsymbol{\Sigma}_R\sigma_B^2, \nonumber \\ 
\boldsymbol{\delta}_{A}^T\boldsymbol{\delta}_{A}&=\boldsymbol{\Sigma}_R \big(\boldsymbol{\delta}_{A}\boldsymbol{\delta}_{A} ^T \big)/P, \\
\boldsymbol{\delta}_{B}^T\boldsymbol{\delta}_{B}&=\boldsymbol{\Sigma}_R \big(\boldsymbol{\delta}_{B}\boldsymbol{\delta}_{B}^T  \big)/P,  \nonumber 
\end{align}

\noindent wherefrom we can obtain the following result: 	

\begin{align}\label{eqSpatialCov6}
{\rm var}\big( \hat{\boldsymbol{\delta}}_{A}\hat{\boldsymbol{\delta}}_{B}^T \big)&= {\rm tr}\Big(\frac{\boldsymbol{\delta}_{B}^T\boldsymbol{\delta}_{B}\boldsymbol{\Sigma}_{R}\boldsymbol{\Sigma}_R\sigma_A^2}{P}+ \frac{\boldsymbol{\delta}_{A}^T\boldsymbol{\delta}_{A}\boldsymbol{\Sigma}_{R}\boldsymbol{\Sigma}_R\sigma_B^2}{P}+ \sigma_A^2\boldsymbol{\Sigma}_R\boldsymbol{\Sigma}_R\sigma_B^2 \Big)
   \nonumber \\
&=\Big( \boldsymbol{\delta}_{B}^T\boldsymbol{\delta}_{B}\sigma_A^2+ \boldsymbol{\delta}_{A}^T\boldsymbol{\delta}_{A}\sigma_B^2+ P \sigma_A^2\sigma_B^2  \Big)  \frac{{\rm tr}\big(\boldsymbol{\Sigma}_{R}\boldsymbol{\Sigma}_R\big)}{P},
\end{align}	

\noindent which equals Eq. \ref{eqProdVar2} for the case of independent voxels, but this time scaled by the factor  ${\rm tr} \big(\boldsymbol{\Sigma}_{R}\boldsymbol{\Sigma}_R\big)/P$. When all voxels are independent, i.e.,  $\boldsymbol{\Sigma}_{P}=I$, ${\rm tr} \big(\boldsymbol{\Sigma}_{R}\boldsymbol{\Sigma}_R\big)=P$ and the correction factor vanishes. For the case of remaining spatial correlations in the data, one has ${\rm tr} \big(\boldsymbol{\Sigma}_{R}\boldsymbol{\Sigma}_R\big)>P$, and the predicted variance will increase. Note that ${\rm tr} \big(\boldsymbol{\Sigma}_{R}\boldsymbol{\Sigma}_R\big)=\sum_{i}\lambda_i^2$ where the 
$\lambda_i$'s  denote the eigenvalues of $\boldsymbol{\Sigma}_R$. Thus our ability to estimate the variability of the LDC correctly will depend on how accurately this quantity can be determined from sample estimates. We will address this problem in section 4.2. 

\subsection{Averaging distances across different cross-validation folds}
In the last step, we need to take into account the averaging of the estimated distances across the $M$ different cross-validation folds. While data from different partitions can be assumed to be independent, the inner products across cross-validation folds are not. This is because the partitions from one cross-validation fold will be again included in other folds. 
The two pattern differences that enter the product in Eq. \ref{eqCrossDist} come from a single partition (that is, $A = m$), or from all other partitions (that is, $B$ corresponding to $\sim m$). As a shorthand for the covariance between difference estimates $i$  and $j$ that are based on partitions $A$ and $B$, we introduce the symbol

\begin{eqnarray}\label{eqAvrg1}
{\Xi}_{i,j}^{A,B}={\rm Cov}\big( \hat{\boldsymbol{\delta}}_{i,A},\,\hat{\boldsymbol{\delta}}_{j,B} \big).
\end{eqnarray}

\noindent This is the covariance for each individual voxel. We now exploit the bilinearity of the covariance operator, that is,

\begin{eqnarray}\label{eqAvrg2}
{\rm Cov}\Big( \sum_{m} {x}_m , \, \sum_{n} {y}_n \Big)=\sum_{m} \sum_{n} {\rm Cov}\big( {x}_m,\,{y}_n \big),
\end{eqnarray}

\noindent to obtain the following general result: 

\begin{align}\label{eqAvrg3}
{\rm Cov}(\hat{d}_i,\,\hat{d}_j)
=&\sum_{m} \sum_{n} \frac{{\rm Cov}\big( \hat{\boldsymbol{\delta}}_{i,m}\hat{\boldsymbol{\delta}}_{i,\sim m}^T,\, \hat{\boldsymbol{\delta}}_{j,n}\hat{\boldsymbol{\delta}}_{j,\sim n}^T  \big)}{M^2P^2}   \nonumber \\
=&\frac{{\rm tr}\big(\boldsymbol{\Sigma}_{R}\boldsymbol{\Sigma}_R\big)}{M^2P^2} 
\sum_{m} \sum_{n} \Big\{ \frac{\boldsymbol{\delta}_i\boldsymbol{\delta}_j^T}{P} \Big({\Xi}_{i,j}^{\sim m\sim n}+{\Xi}_{i,j}^{\sim m n}+{\Xi}_{i,j}^{ m\sim n}+{\Xi}_{i,j}^{ m n}\Big)  \nonumber \\
&+ \Big({\Xi}_{i,j}^{ m n}{\Xi}_{i,j}^{\sim m\sim n}+{\Xi}_{i,j}^{ m\sim n}{\Xi}_{i,j}^{\sim m n} \Big)\,\Big\}   \nonumber \\
=&\frac{{\rm tr}\big(\boldsymbol{\Sigma}_{R}\boldsymbol{\Sigma}_R\big)}{P^2} \Big\{ \frac{\boldsymbol{\delta}_i \boldsymbol{\delta}_j^T}{P}   {S}_{i,j}+{N}_{i,j} \Big\} 
\end{align}

\noindent where

\begin{eqnarray*}
	{S}_{i,j}=\frac{1}{M^2}\sum_{m} \sum_{n}  \Big\{ {\Xi}_{i,j}^{\sim m\sim n}+{\Xi}_{i,j}^{\sim m n}+{\Xi}_{i,j}^{ m\sim n}+{\Xi}_{i,j}^{ m n} \Big\} 
\end{eqnarray*}
\noindent and

\begin{eqnarray*}
	{N}_{i,j}=\frac{1}{M^2}\sum_{m} \sum_{n}  \Big\{ {\Xi}_{i,j}^{ m n}{\Xi}_{i,j}^{\sim m\sim n}+{\Xi}_{i,j}^{ m\sim n}{\Xi}_{i,j}^{\sim m n} \Big\}.   \nonumber
\end{eqnarray*}

\noindent This is the most general expression of the LDC variance, which can even be used when the covariance structure of different partitions ($\boldsymbol{\Sigma}_K$) differs from each other (section 3.5). For the basic case whereby the difference estimates from all $M$ partitions can be assumed to have the same covariance, that is, $\mathbf{\Xi}_{i,j}\equiv {\rm Cov}(\hat{\boldsymbol{\delta}}_{i,m},\,\hat{\boldsymbol{\delta}}_{j,m})$, a dramatic simplification can be obtained. In this instance the best estimate of $\boldsymbol{\delta}_{\sim m}$ is the average of all partitions except $m$: 

\begin{eqnarray}\label{eqAvrg4}
\hat{\boldsymbol{\delta}}_{\sim m}=\sum_{n\neq m} \hat{\boldsymbol{\delta}}_n/(M-1).
\end{eqnarray}

\noindent Accordingly, we have

\begin{eqnarray}\label{eqAvrg5}
\boldsymbol{\Xi}^{A,B} = \threepartdef
{\boldsymbol{\Xi}}      {A=m,\,B=m}
{\mathbf{0}}      {A=m,\,B=\sim m}
{\boldsymbol{\Xi}/(M-1)} {A=\sim m,\,B=\sim m,}
\end{eqnarray}

\noindent and for $n\neq m$,

\begin{eqnarray}\label{eqAvrg6}
\boldsymbol{\Xi}^{A,B} = \threepartdef
{\mathbf{0}}      {A=m,\,B=n}
{\boldsymbol{\Xi}/(M-1)}      {A=m,\,B=\sim n}
{(M-1)\boldsymbol{\Xi}/(M-1)^2} {A=\sim m,\,B=\sim n.}
\end{eqnarray}

\noindent Substituting  the elements of the appropriate  representations of $\boldsymbol{\Xi}^{A,B}$ into Eq. \ref{eqAvrg3} and summing up, we have

\begin{align}\label{eqAvrg7}
{S}_{i,j}=\ &\frac{1}{M^2}\Big\{ M\big( \frac{{\Xi}_{i,j}}{M-1} + {\Xi}_{i,j} \big)\nonumber \\
&+M(M-1) \Big( \frac{(M-2){\Xi}_{i,j}}{(M-1)^2}+\frac{2{\Xi}_{i,j}}{M-1} \Big)\,\, \Big\} \nonumber \\
=\ &\frac{1}{M^2}\,{\Xi}_{i,j} \Big\{ \frac{M}{M-1}+M+\frac{M(M-2)}{M-1}+2M \Big\} \nonumber \\
=\ &\frac{4}{M}\,{\Xi}_{i,j},  \nonumber \\
{N}_{i,j}=\ &\Big\{ \frac{M{\Xi}_{i,j}{\Xi}_{i,j}}{M-1}+\frac{M(M-1){\Xi}_{i,j}{\Xi}_{i,j}}{(M-1)^2}  \Big\}  \\
=\ &\frac{2\,{\Xi}_{i,j}{\Xi}_{i,j}}{M(M-1)}, \nonumber
\end{align}

\noindent and 

\begin{eqnarray*}
{\rm Cov}(\hat{d}_i,\,\hat{d}_j)=\frac{{\rm tr}\big(\boldsymbol{\Sigma}_{R}\boldsymbol{\Sigma}_R \big)}{P^2} 
\Big( 4\frac{\boldsymbol{\delta}_{i}\boldsymbol{\delta}_{j}^{T}}{P}
\frac{\Xi_{i,j}}{M}+2\frac{{\Xi}_{i,j}{\Xi}_{i,j}}{M(M-1)}\Big). \nonumber
\end{eqnarray*}

\noindent Finally, on writing the desired complete covariance matrix using element-by-element multiplication, we obtain the result given in Eq. \ref{eqApproxVar}.

\subsection{Estimation of the condition-by-condition covariance $\boldsymbol{\Sigma_K}$}

If the condition-by-condition covariance of the activity estimates $\boldsymbol{\Sigma_K}$ can be assumed to be equal across the different partitions, then the best estimates for the activity differences is Eq. \ref{eqAvrg4} and the variance of the LDC is given by Eq. \ref{eqApproxVar}. In this case the condition-by-condition covariance can be estimated directly from the prewhitened activity estimates, averaged over the voxels: 

\begin{eqnarray}\label{eqSimSigmaK}
\mathbf{\hat{\boldsymbol{\Sigma}}}_{K} = \sum_{m} (\mathbf{\hat{U}}_m - \mathbf{\bar{U}}) (\mathbf{\hat{U}}_m - \mathbf{\bar{U}})^{T}/(M-1)P.
\end{eqnarray}

\noindent The more general result specified by Eq. \ref{eqAvrg3} is useful when the covariance structure of the activation estimates ($\mathbf{B}_m$) cannot be assumed to be constant across partitions. In neuroimaging experiments this can easily occur when different imaging runs have different sequences of conditions (inducing different correlations between the estimates), or if the number of conditions per run is not balanced (for example due to the exclusion of error trials).  When  the structure of the estimation is not the same across cross-validation folds, the best estimate of the pattern difference is different from that given in Eq. \ref{eqAvrg4} and depends on the structure of the design matrix of the partitions. If set $A$ includes all partitions except partition $m$, then the best linear estimate of a pattern distance is given by 

\begin{eqnarray}\label{eqEst3}
\hat{\boldsymbol{\delta}}_{i,A} = \mathbf{c}_{i} \big( \mathbf{X}^T_A \hat{\boldsymbol{\Sigma}}_T^{-1} \mathbf{X}_A \big)^{-1} \mathbf{X}^T_A \hat{\boldsymbol{\Sigma}}_T^{-1} \mathbf{Y}_{A},
\end{eqnarray}

\noindent where the design matrix $\mathbf{X}_A$ contains the concatenated design matrices from all included partitions, and $\hat{\boldsymbol{\Sigma}}_{T}$ is an appropriately sized estimate of the temporal autocovariance matrix. It follows that

\begin{eqnarray}\label{eqEst4}
{\Xi}_{i,j}^{AB}=\mathbf{c}_{i}\big( \mathbf{X}_A^T \hat{\boldsymbol{\Sigma}}_T^{-1} \mathbf{X}_A \big)^{-1} \mathbf{X}_A^T \mathbf{X}_B \big( \mathbf{X}_B^T \hat{\boldsymbol{\Sigma}}_T^{-1} \mathbf{X}_B \big)^{-1}\mathbf{c}_{j}^T.
\end{eqnarray}

\noindent Note that Eq. \ref{eqSimSigmaK} has an added advantage over Eq. \ref{eqEst4} as it does not depend on a valid estimate of the temporal autocorrelation. The latter can be difficult to obtain using current standard analysis packages \cite{RN3550}. Thus, in general, the assumption of equal partitions and making use of Eq. \ref{eqSimSigmaK} may be recommended.  

\section{Numerical simulations}

\subsection{Normal approximation with low number of voxels}
This section presents numerical simulations to test the accuracy of the suggested approximation to the sample distribution of the LDC. We start with a simulation, in which we generate $\mathbf{\hat{U}}$ directly from a known multivariate normal distribution. We used $K=5$ conditions and $M=5$ runs. To assess the limits of the normal approximation, we used $P=30$ voxels, which is a comparatively low number for fMRI analyses. We first generated data assuming that all the true distances are equal to zero.  We then varied the signal level $\sigma^{2}_{a}$ from 0.05 to 0.2, setting the true distance between conditions 1 and 2 to $d=1.5\sigma^{2}_{a}$, the distances between conditions 1 and 3-5 to $d=1\sigma^{2}_{a}$, and the distances between conditions 2 and 3-5 to $d=0.5\sigma^{2}_{a}$. In this simulation, both the noise and the signal were spatially independent across voxels.

	\begin{figure}
		\centering
		\includegraphics[width=9cm]{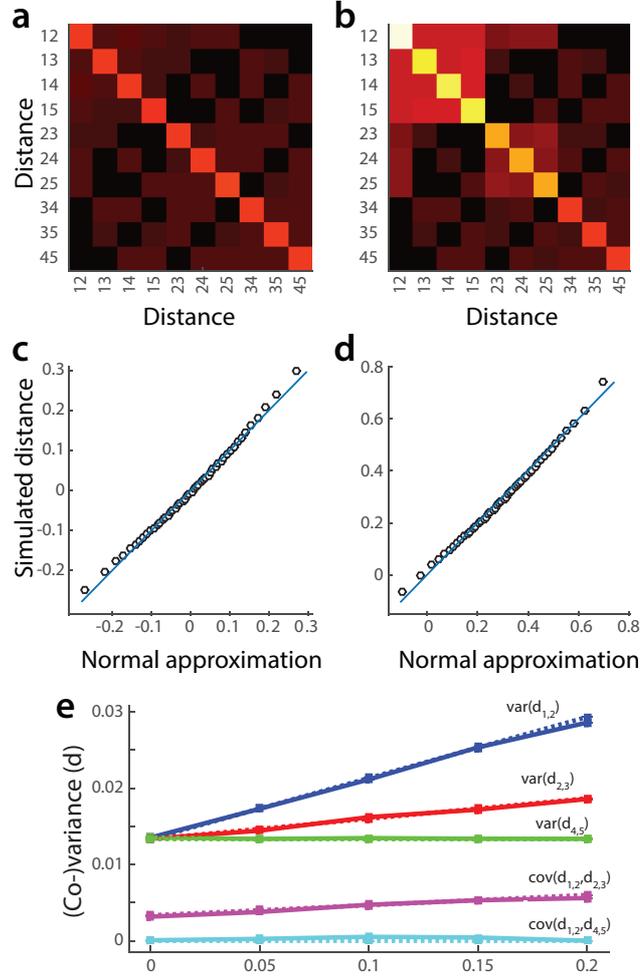}
		\caption{Joint sample distribution of the distance estimates (LDC) between 5 conditions. 
			(\textbf{a}) Covariance matrix between distance estimates when the true distances $d=0$. 
			(\textbf{b}) Covariance matrix with $d_{1,2}=0.3$, $d_{1,3-5}=0.2$, $d_{2,3-5}=0.1$. 
			(\textbf{c}) Q-Q plot of the estimated distances and the normal approximation for zero distances. 
			(\textbf{d}) Q-Q plot of the estimated distances and the normal approximation for zero distances.
			(\textbf{e}) Covariances of the distances as the true distances increase. The dotted line shows the numerical prediction.}
		\label{fig2}
	\end{figure}

Fig. 2a shows the variance-covariance matrix of the 10 distances for $d=0$. As expected, distances between no-overlapping sets of conditions (i.e. $d_{1,2}$ and $d_{4,5}$) are uncorrelated, whereas distances that share one condition correlate positively. As the true distances increase (Fig. 2b), their variance increases with the mean - the largest variance being measured for the largest distance ($d_{1,2}$). Furthermore, the covariance between non-zero distances ($d_{1,2}$, $d_{2,3}$) also increases. The numerical expression for the expected variance (Eq. \ref{eqApproxVar}) accurately predicts these increases in variance and covariance with increasing mean (Fig. 2e). 

Fig. 2c,d shows quantile-quantile plots for the empirical distribution of the LDC against the predicted normal approximation (Eq. \ref{eqApproxNorm}). Even with low numbers of voxels, the distribution is well approximated, although the tails show some deviation from normality. The left tail of the distribution is slightly lighter than predicted, the right tail is heavier. This deviation may result in increased Type I errors when using the normal approximation for statistical tests. However, in typical analyses, we use $P>60$ voxels, for which the normal approximation of the tails becomes increasingly accurate. 

\subsection{Spatially dependent noise and prewhitening}
In the previous section we have used spatially independent data. In reality, however, we need to rely on prewhitening to make the data more independent, and we need to estimate $\boldsymbol{\Sigma}_{R}$ in order to utilize Eq. \ref{eqApproxVar}. 

Accordingly, in this section, we verify the determination of the variance of LDC under more realistic conditions. We based our simulation on parameter values derived from a real fMRI experiment, in which we mapped the finger representations in primary motor and sensory cortex \cite{RN3031, RN3415}. In this study, we employed an event-related design in which participants tapped one of their 10 fingers for $8.1s$. Hence, there were $K=10$ conditions, and each condition was measured 3 times during each of the $M=8$ independent imaging runs or partitions. Each run contained $T_{m}=123$ images (time points) with a sampling rate of $2s$. To simulate the temporal dependence of fMRI noise \cite{RN2203, RN2073}, we approximated the temporal autocorrelation observed in this experiment using a double exponential function: 

\begin{eqnarray}\label{eqSimTemp}
{\rm r}(t,t+\tau)=0.5 {\rm exp}(-\tau) + 0.5 {\rm exp}(-\tau/40).
\end{eqnarray}

\noindent We generated data for a spherical ROI with a radius of $8mm$, assuming a voxel size of $2 \times 2 \times 2mm$, resulting in $P=257$ voxels. The spatial correlation between two voxels depended on their spatial distance $p_{1}-p_{2}$, using an exponential kernel of width $s_{\epsilon}$, with 

\begin{eqnarray}\label{eqSimSpace}
{\rm r}(1,2)={\rm exp}(-(p_{1}-p_{2})^2/s^{2}_{\epsilon}).
\end{eqnarray}

The simulated noise data was analyzed separately for each run, using a design matrix that included an intercept and one regressor per condition. Each regressor consisted of a boxcar function covering the duration of the 3 trials. As is typical for fMRI analysis, the boxcar function was convolved with a standard estimate of the hemodynamic response function \cite{RN2030} to arrive at a predicted overall response. 
We estimated the activity patterns and the spatial autocorrelation of the noise using Eq. \ref{eqEstBeta} and \ref{eqWhiteSigmaP}, and prewhitened the activity estimates (Eq. \ref{eqWhitePrewhite}) using a regularised estimate of the spatial covariance matrix (Eq. \ref{eqWhiteRegular}). These estimates were then used to determine the LDC distances between the 10 conditions. We systematically varied the regularization constant between $h=0.1$ and $1$. 

The variability of these simulated distances for different widths of the spatial noise kernel (x-axis) and for different values of the regularisation constant (different plots) are shown as solid lines in Fig. 3. 

To predict the size of the variance from a single data set, we either assumed that the voxels were independent after prewhitening ($\mathbf{\boldsymbol{\Sigma}}_{R}=\mathbf{I}$, Fig. 3, gray line) or we used the estimate from Eq. \ref{eqWhiteSigmaR} for the residual spatial correlation (Fig. 3, dashed line). 

The actual variance was systematically larger than the predicted variability assuming independent voxels (gray line), indicating that the assumption of spatial independence after prewhitening is too optimistic. When using an estimate for the residual variance (dashed line), the situation depended on the size of the shrinkage coefficient. When the estimate was based only on the diagonal of the covariance matrix ($h=1$), we systematically overpredicted the variance, due to the fact that the expected value of ${\rm tr}(\hat{\boldsymbol{\Sigma}}_{R}\hat{\boldsymbol{\Sigma}}_{R})$ is larger than ${\rm tr}(\boldsymbol{\Sigma}_{R}\boldsymbol{\Sigma}_{R})$. For a shrinkage coefficient of $h=0.4$, the prediction was accurate for all tested widths of the noise kernel. Coincidentally, this is also well in the range of values for which the split-half reliability of the distance estimates using the empirical data from \cite{RN3031, RN3415} becomes maximal (see Fig. 4). For $h=0.2$, we under-predicted the variance for spatially independent data. However, for spatial noise correlations with realistic values ($\sigma_{\epsilon}>2.5$), the variance estimate approached the simulated value. 

In sum, these results show that the proposed normal approximation is sufficiently accurate for the parameter range typically encountered in fMRI experiments. Furthermore, a regularization constant of approximately $0.4$ is recommended to further improve the estimation accuracy of the variance. 

	\begin{figure}
		\centering
		\includegraphics[width=9cm]{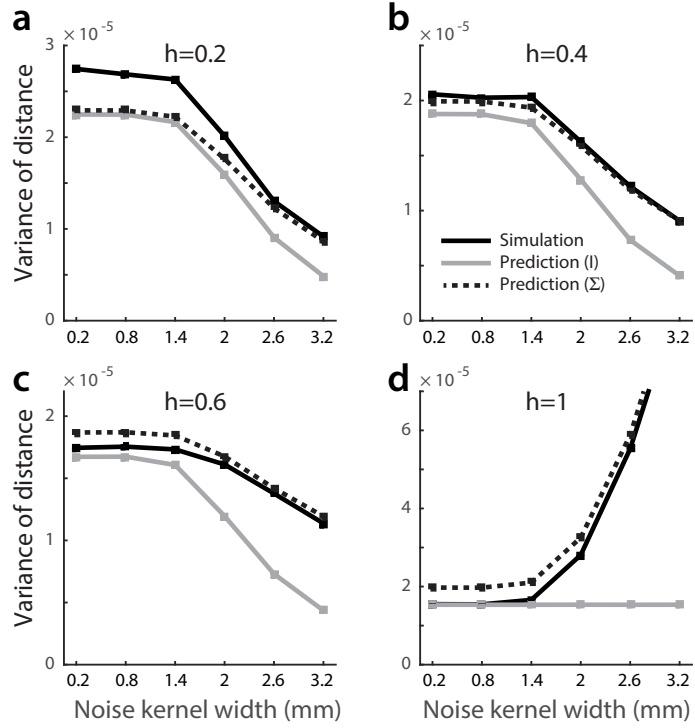}
		\caption{Simulated and predicted variance of LDC distances as a function of the width of the spatial noise kernel ($\sigma_{\epsilon}$). The variances of the simulated distances are shown in the black solid line. Distances are calculated on prewhitened activity estimates, using regularised estimates of the spatial noise covariance with regularisation factors of (\textbf{a}) $h=0.2$, (\textbf{b}) $h=0.4$, (\textbf{c}) $h=0.6$, (\textbf{d}) $h=1$. The predicted variance assuming spatially independent voxels after prewhitening is shown in gray; the predicted variance using an estimated residual spatial correlation (Eq. \ref{eqWhiteSigmaR}) is shown as a dashed line.}
		\label{fig3}
	\end{figure}

\begin{figure} 
	\centering
	\includegraphics[width=6cm]{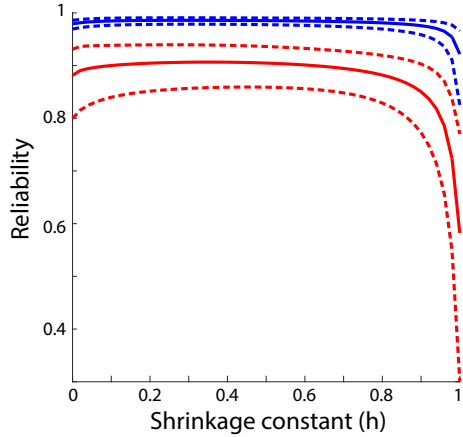}
	\caption{Average split-half reliabilities of pattern distance estimates in primary motor cortex for finger movements of the contralateral (blue) and ipsilateral (red) hands \cite{RN3243}. Activity patterns were prewhitened using different values of $h$ for Eq. \ref{eqWhiteRegular} with 0 corresponding to no shrinkage and 1 to using only the diagnonal of the spatial covariance matrix. The $10$ distances for each hand were then calculated separately for odd and even and finally correlated. The dotted lines indicate the 95\% confidence interval.}
	\label{fig4}
\end{figure}

\section{Applications}
\subsection{Significance test for individual distances}
The normal approximation to the LDC distance enables us to obtain confidence intervals on distance estimates, and to generate significance tests as to whether a specific distance (or a linear combination of distances) is significantly larger than zero. Such a test is important in the context of RSA as it indicates that a given area encodes the difference between two conditions. 

For illustration we used the same experiment as in the last section, however this time generating data for a real region-of-interest (ROI) of $P=375$ voxels with a spatial covariance matrix estimated from the original experiment. 
We simulated 10,000 replications of the experiment, each time estimating the 45 distances between the 10 fingers. For each simulation, we used the estimates for $\boldsymbol{\Sigma}_{K}$ and $\boldsymbol{\Sigma}_{R}$, to obtain a predicted variance-covariance matrix $\mathbf{V}$. To test whether a specific linear combination $\mathbf{c}$ of the estimated distances is zero, one can define the following z-test statistic: 

\begin{eqnarray}\label{eqAppTest}
z = \frac{\mathbf{c}^{T} \mathbf{\hat{d}}}{\sqrt{\mathbf{c^{T} V c}}}.
\end{eqnarray}

To test whether an individual distance is larger than zero, $\mathbf{c}$ would be a vector of zeros with a single 1 at the place of the distance to be tested. For the average distance, one would use a vector of ones, etc. The false positive rate, i.e., the number of $z>\Phi(1-\alpha)$ was $0.0497$ for $\alpha=0.05$ and $0.0122$ for $\alpha=0.01$, again demonstrating the appropriateness of the normal approximation in a realistic scenario. 

Applied to the real data from the left primary motor cortex in the a single subject from \cite{RN3243}, we obtained a significant z-value for the average distance between contralateral (right) fingers of $z=21.08$, but also a significant result for the fingers of the ipsilateral (left) hand of $z=14.58$. Whereas all individual contralateral finger pairs differed significantly from each other, only 6 of the 10 ipsilateral finger pairs were significant at a $p<0.05$ level. 

The z-statistic (Eq. \ref{eqAppTest}) can also be applied to tests on the difference between two distances; for example one can test the hypothesis that the distance $d_{1,2}$ is smaller than the distance $d_{1,5}$. In this case $\mathbf{V}$ needs to be predicted under the null hypothesis that $d_{1,2} = d_{1,5}$, and not, as before, under the hypothesis that $d_{1,2} = d_{1,5} = 0$. In a simulation in which the true distances between all fingers were set to $d=0.01$, the latter test would result in a false-positive rate of $p=0.09$ for $\alpha=0.05$. Using the mean of $\hat{d}_{1,2}$ and $\hat{d}_{1,5}$ as the assumed true value in the calculation of $\mathbf{V}$ reduced this value to the $p=0.05$ mark. 

In sum, the normal approximation provides a simple and straightforward way of testing linear combinations of distances, freeing the user of the need to perform permutation statistics \cite{RN2730, RN3312}. The significance test allows one, for example, to carry out efficient statistical thresholding of representational dissimilarity matrices (e.g., see \cite{RN3300}), and to test for differences between distances on the single-subject level. 

\subsection{Model comparison}
A second important application of the normal approximation lies in the comparison of different representational models. We define representational models as explicit hypotheses that predict the relative distances between conditions, i.e., how different the neural codes for different conditions are from one another. The power of this approach lies in the fact that one does not have to worry about which voxel corresponds to which unit of the model, but only whether the overall representational structures match. 

Consider the two models of movement representations in the primary motor cortex shown in Fig 5. The first model (Fig. 5a,b) states that the representational structure is determined by the covariances between movements of the fingers in everyday life \cite{RN2703}. The second model (Fig. 5c,d) predicts that the representational structure is determined by the similarity of muscle activities required to produce the patterns. While the prediction of the two models have some commonalities (e.g., the distances involving the thumb, first row, are the largest), the two models make distinguishable predictions for an experiment in which five movements of the individual fingers were tested (Fig. 5a,c), and an experiment in which $31$ different hand movements were tested \cite{RN3415} (Fig. 5b,d). 

	\begin{figure}
		\centering
		\includegraphics[width=7cm]{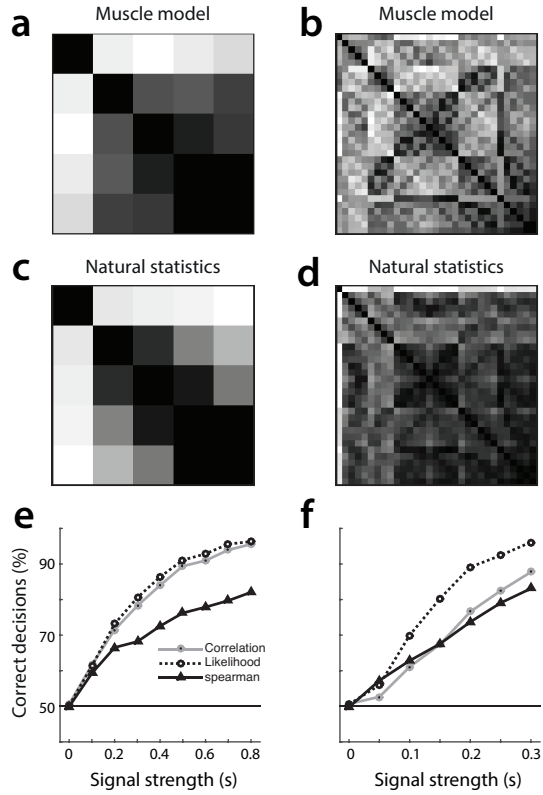}
		\caption{Comparing representational models. 
			(\textbf{a}) Predicted distances (black 0 - white 1) for the 5 fingers from the natural statistics of movement. 
			(\textbf{b}) Predictions from the same model for an experiment with 31 different multi-finger movement conditions. 
			(\textbf{c}) Predicted distances for the 5 fingers of the contralateral hand from the pattern of muscle activities. 
			(\textbf{d}) Predictions from the same model for an experiment with 31 movement conditions. 
			(\textbf{e}) Percentage of correct model selection decisions on simulated data with varying degrees of signal strength ($s$). The models were selected on the basis of rank correlations (black solid line), cosine angles (gray solid line) and approximated likelihoods (Eq. \ref{eqAppLogLike}, dashed line).
			(\textbf{f}) Model selection accuracy for the experiment with 31 conditions.}
		\label{fig5}
	\end{figure}

The models make only predictions about the relationship or ratios between the distances; the absolute size of the distances depends on the signal-to-noise level of the fMRI signal for that particular subject and region. That is, the expected distance under a representational model is 

\begin{eqnarray}\label{eqAppModel}
 \mathbf{d} = s \mathbf{m}, 
\end{eqnarray}

\noindent where $\mathbf{m}$ is a $D \times 1$ vector of predicted distances from a particular model and $s$ an arbitrary (but positive) scaling factor for a particular subject. 

Traditionally, RSA models have been evaluated using the relatively cautious approach of comparing the rank-ordering of the $D$ predicted distances with the rank-ordering of the estimated distances \cite{RN3300}. The use of the LDC distance, however, makes the zero-point of the distances meaningful (implying that the activity patterns for these two conditions are identical), a fact that also renders the ratio of distances interpretable. Thus, an alternative approach to rank-based correlations is to compute the cosine angle between the model predictions $\mathbf{m}$ and the estimated distances 

\begin{eqnarray}\label{eqAppCosineCorr}
r = \mathbf{m}^{T} \mathbf{\hat{d}} / \sqrt{\mathbf{m}^{T} \mathbf{m} \,\mathbf{\hat{d}}^{T} \mathbf{\hat{d}}}.
\end{eqnarray}

Ultimately, however, this expression will be very sensitive to large predicted distances, as these also have much larger leverage than smaller distances. Unfortunately, large distances are also estimated with a higher variance. Using the normal approximation (Eq. \ref{eqApproxNorm}), we express the log-likelihood of the distances under the model as

\begin{eqnarray}\label{eqAppLogLike}
{\rm log}\,{\rm p}(\mathbf{\hat{d}} | \mathbf{m}, s) = 
-\frac{D}{2}{\rm log}(2\pi)-\frac{1}{2}{\rm log}(|\mathbf{V}|)-\frac{1}{2}(\mathbf{\hat{d}} - \mathbf{m}s)^{T}\,\mathbf{V}^{-1}\,(\mathbf{\hat{d}} - \mathbf{m}s).
\end{eqnarray}

Note that $\mathbf{V}$ not only depends on $\boldsymbol{\Sigma}_{K}$ and $\boldsymbol{\Sigma}_{R}$, which are estimated, but also on the model prediction $\mathbf{m}$ and the scaling factor $s$. The latter can be estimated by maximizing Eq. \ref{eqAppLogLike}, using iteratively-reweighted least squares. Starting from an arbitrary value for $\hat{s}$, one estimates $\mathbf{V}$ and then obtains an updated estimate 

\begin{eqnarray}\label{eqAppIRLS}
\hat{s}^{*} = \big(\mathbf{m}^{T}\mathbf{V}^{-1}\mathbf{m})^{-1}\,\mathbf{m}^{T} \mathbf{V}^{-1}\mathbf{\hat{d}},
\end{eqnarray}

\noindent iterating until convergence is obtained. 

 For the two experiments, we simulated distances according to the two competing models. For each simulation we then evaluated the two competing models using Spearman rank correlation, the cosine angle (Eq. \ref{eqAppCosineCorr}) or the approximated likelihood (Eq. \ref{eqAppLogLike}) using the maximum-likelihood estimate for the scaling factor. The model with the higher criterion value was selected and the correct model selection decisions were counted. 
 
 The percentage of correct decisions increased with increasing signal strength (Fig. 5e,f). For the 5-condition experiment, the cosine angle and likelihood approach clearly outperformed the rank correlation approach. This is expected, as the rank ordering of the distances was relatively similar across the two models, with the differences mostly residing in the predicted ratios between distances. For the 31-condition experiment, however, the likelihood approach performed better than the other two criteria, likely because both models contained a few large distances that dominated the cosine angle. 
 
 In sum, the likelihood-based approach provides an elegant and powerful way to compare representational models. As an extension of this example, we can also consider models with more free parameters. These can also be estimated using an iteratively-reweighted least squares approach as outlined in Eq. \ref{eqAppIRLS}. In general, the normal approximation allows users to make use of the large toolkit of regression techniques, including Bayesian regression and model comparison using marginal likelihoods. 

\section{Discussion}
In this paper, we derived and tested a normal approximation to the sampling distribution of the cross-validated Mahalanobis distance or linear discriminant contrast (LDC). The LDC serves as a measure of dissimilarity to quantify the relationship between neural activity patterns. It is central to RSA, which is a powerful analysis approach with which on can relate the structure of neuronal activity patterns across a variety of measurement techniques and compare these to theoretical predictions from computational models \cite{RN2697, RN3225, RN3544}. In this context, the LDC has a number of advantageous features: Due to its multivariate noise normalization, it tends to be more reliable than other distance measures \cite{RN3540}. The cross-validated nature of the quadratic form makes the measure unbiased - i.e., the expected mean of the sampling distribution equal the true distance between the two patterns. Therefore, the measure will be on average zero if the true patterns are identical, making it interpretable across different levels of measurement noise. This feature is especially important when comparing the representational structure across different individuals, brain regions, or techniques, all of which can differ in their amount of measurement noise. 

The main advance introduced in this paper is the derivation of an analytical expression for the covariance matrix of a sample LDCs. The expression has two parts: one component only depends on the variability of the underlying activity estimates; the other is signal-dependent and increases as the true distances increase. The second component arises from the fact that the true signal is multiplied by the noise in the quadratic form. In sum, the variance of the LDC increases linearly in the true distance with a non-zero offset when the true distance is zero. 

This feature also prevents the use of a simple non-linear transform of the LDC, which would render the variance constant. For example, in the analysis of Poisson-distributed neuronal spike rates, the square-root transform is widely used \cite{RN3552}, as it renders the variance of the resultant measure approximately independent of the mean. However, due to the non-zero offset of the variance (which can be substantial), such simple transform is not available for the LDC. 

We show that the normal distribution using analytical expressions for mean and variance provides a sufficiently accurate approximation to the sampling distribution of the LDC. The goodness of the approximation hinges on two factors: First, the number of voxels should be higher than 30 to guarantee that the tails of the distribution are fitted well. This, however, is usually given, as multivariate analyses are typically performed in regions of interest or searchlights with $P>50$ voxels. Secondly, the multivariate noise estimate used for prewhitening should be regularized using a factor of $0.2-0.5$ to ensure good accuracy of the variance approximation. This range of regularization constants is recommended, as it also maximizes the reliability of the resultant distance measures \cite{RN3540}. 

While the normal approximation has universal utility for a whole range of secondary analyses based on the LDC, we have focussed here on two examples: First, it allows a straightforward test to assess the significance of any linear contrast of LDCs from the same sample. Currently such inferences in the context of multivariate analysis are made using permutation tests \cite{RN3312, RN3316}, in which the labels for the conditions within each imaging runs are exchanged. Apart from the added computational costs, permutation tests are typically less powerful than good closed-form approximations. Furthermore, while it is straightforward to define an exchangeability condition for the hypothesis that two conditions have the same pattern (i.e. that a distance is zero), it is much harder to determine exchangeability conditions for the hypothesis that two distances are the same. 

As an alternative to permutation tests, Nili {\sl{et al.}} \cite{RN3300} proposed the use of an approximately t-distributed transform of the LDC. This approximation, however, has a number of distinct disadvantages compared to the approximation used here: it can only be applied to distances computed using a single cross-validation fold, whereas in typical fMRI experiments more independent partitions are used. Furthermore, the expression ignores the covariance between related distance estimates, and the possible signal dependence of the variance. Therefore, the approach can only be used to assess significance of individual distances in a very restricted set of applications. The z-test proposed in this paper retains validity over a larger range of situations. 

The second application example lies in the comparison of two representational models based on a whole matrix of dissimilarities. In this case, the normal approximation is utilized to derive a closed-form expression of the data likelihood. This expression can be used to estimate the free parameters of the model; here simply the single signal strength parameter. In this context, iterative procedures, such as iteratively-reweighted least squares, are necessary to take the signal-dependent variability into account. The likelihood can then serve as a criterion for model selection. As shown in two examples, this approach provides more accurate model selection, as compared to rank correlations \cite{RN3300, RN3544}, Pearson correlations \cite{RN3415}, or cosine angles. 

In this paper we have focused on the application of RSA to fMRI data. RSA is being applied widely to other methodologies in neuroscience, such as invasive neurophysiology and recordings of evoked potentials from the scalp. In all these applications, the cross-validated Mahalanobis distance provides a powerful way to quantify the similarity of multivariate patterns - especially when the signal-to-noise level of individual measurements is low. The normal approximation presented here lays the foundation for secondary analyses based on this measure.

	
\bibliography{bblibrary}
\bibliographystyle{plain}

	\newpage

	\begin{appendices}
		\section{Summary of Notations}

		\begin{flushleft}
			\begin{tabular}{ l l l }
			\textbf{Symbol} &\textbf{Dimension}& \textbf{Meaning} \\ 
			& &\\
				$T$ & $1$ & Overall number of time points in the data set \\ 
				$T_m$ & $1$ & Number of time points in partition $m$  \\ 				
				$M$ & $1$ & Number of independent partitions of the data (imaging runs) \\ 				
				$K$ & $1$ & Number of conditions (stimuli) \\ 				
				$P$ & $1$ & Number of measurement channels (voxels in fMRI)   \\ 				
				$D$ & $1$ & Number of pairwise distances, usually $K(K-1)/2$ \\ 				
				$\mathbf{B}$ & $K\times P$ & Matrix of true activation patterns \\ 				
				$\hat{\mathbf{B}}_{m}$ & $K\times P$ & Matrix of estimated activation patterns from partition $m$ \\ 				
				$\mathbf{U}$ & $K\times P$ & Matrix of true whitened activation patterns \\ 				
				$\hat{\mathbf{U}}_{m}$ & $K\times P$ & Matrix of estimated whitened activation patterns from partition $m$ \\ 				
				$\hat{\mathbf{u}}_{i,m}$ & $1\times P$ & Activation pattern for ith condition from partition $m$ \\ 				
				$\mathbf{X}_m$ & $T_m\times K$ & Design matrix to estimate activation from time series  \\  
							
				$\mathbf{Y}_m$ & $T_m\times P$ & Raw time series data from partition $m$\\ 				
				$\mathbf{R}_m$ & $T_m\times P$ & Residuals from first-level regression model for partition $m$   \\ 				
				$\boldsymbol{\delta}_j$ & $1\times P$ & Pattern difference between the $j^{\rm th}$ pair of conditions  \\ 				
				$\hat{\boldsymbol{\delta}}_{j,m}$ & $1\times P$ & Estimate of the pattern difference from partition $m$ \\ 				
				$\mathbf{C}$ & $D\times K$ & Contrast matrix, each of the rows defining \\
				& &  one pairwise contrast between conditions\\ 				
				$d_j$ & $1$ & True value of the $j^{\rm th}$ distance, defined between condition $i$ and $k$ \\ 				
				$\hat{d}_j$ & $1$ & Estimated $j^{\rm th}$ distance, integrated over partitions \\ 				
				$\mathbf{d}$ & $D\times 1$ & Vector of all pair-wise true distances \\ 				
				$\mathbf{V}$ & $D\times D$ & Covariance matrix of estimates of $\mathbf{d}$ \\ 				
				$\boldsymbol{\Sigma}_P$ & $P\times P$ & Spatial covariance of  noise across the $P$ voxels  \\ 				
				$\boldsymbol{\Sigma}_R$ & $P\times P$ & Residual voxel-by-voxel covariance matrix after prewhitening   \\ 								
				$\boldsymbol{\Sigma}_T$ & $T\times T$ & Temporal covariance of time-series data \\ 				
				$\boldsymbol{\Sigma}_K$ & $K\times K$ & Covariance matrix of rows of $\hat{\mathbf{U}}_m$  \\ 				
				$\boldsymbol{\Xi}$ & $D\times D$ & Covariance matrix of pattern differences 
			\end{tabular}
		\end{flushleft}
		
	\end{appendices}

\end{document}